\documentclass[12pt,a4paper]{article}
\usepackage{graphicx,multicol}
\usepackage{url}
\usepackage{hyperref,rotating}
\begin{document}
\textwidth=135mm
 \textheight=200mm
\begin{center}
{\bfseries New Bayesian analysis of hybrid EoS constraints with mass-radius
data for compact stars
}
\vskip 5mm
A.~Ayriyan$^{a}$\footnote{Email: ayriyan@jinr.ru}, D.E.~Alvarez-Castillo$^{b,c}$,
D.~Blaschke$^{b,d}$, H.~Grigorian$^{a,e}$,
M.~Sokolowski$^{d}$
\vskip 5mm
{\small {\it $^a$ Laboratory of Information Technologies, JINR Dubna, Russia}}\\
{\small {\it $^b$ Bogoliubov Laboratory of Theoretical Physics, JINR Dubna, Russia}}\\
{\small {\it $^c$ Instituto de F\'{i}sica, Universidad Aut\'{o}noma de San Luis Potos\'{i},
         M\'{e}xico}}\\
{\small {\it $^d$ Institute of Theoretical Physics, University of Wroclaw, Poland}}\\
{\small {\it $^e$ Department of Physics, Yerevan State University, Armenia}}
\\
\end{center}
\vskip 5mm
\centerline{\bf Abstract}
We suggest a new Bayesian analysis using disjunct mass and radius constraints for
extracting probability measures for cold, dense nuclear matter equations of state.
One of the key issues of such an analysis is the question of a deconfinement transition
in compact stars and whether it proceeds as a crossover or rather as a first order
transition.
The latter question is relevant for the possible existence of a critical endpoint in
the QCD phase diagram under scrutiny in present and upcoming heavy-ion collision
experiments.
\vskip 10mm
\section{\label{sec:intro}Introduction}
The most basic features of a neutron star (NS) are the radius $R$ and the mass $M$ which
so far have not been well determined simultaneously for a single object.
In some cases masses are precisely measured like in the case of binary systems, but radii
are quite uncertain \cite{Miller:2013tca}.
In the other hand, for isolated neutron stars some radius and mass measurements exist but
lack the necessary precision to allow conclusions about their interiors.
In fact, it has been conjectured that there exists a unique relation between $M$  and $R$
for all neutron stars and their equation of state (EoS),
thus constraining the properties of their interior \cite{Lindblom1984}.
For this reason, accurate observations of masses and radii are crucial to study cold dense
nuclear matter expected to exist in neutron stars.

However, the presently observational data allow to make only a probabilistic estimations
of the internal structure of the star, via Bayesian analysis (BA).
This technique has been applied for the first time to this problem by Steiner et al.
\cite{Steiner:2010fz}
who exemplified very well the power of this method. In their analysis, however, only a
particular type of objects (X-ray bursters) has been considered under strongly model
dependent assumptions.
In this work is a continuation of our preliminary probabilistic studies of the superdense
stellar matter equation of state using Bayesian Analysis and modeling of relativistic
configurations of neutron stars  \cite{Alvarez-Castillo:2014xea,Blaschke:2014via}
We put special emphasis on the choice of observational constraints and focus on
investigations of the possible existence of deconfined quark matter in massive neutron
stars such as the recently observed $2~M_\odot$ pulsars
\cite{Demorest:2010bx,Antoniadis:2013pzd}.

\section{NS structure and EoS}
The microscopical properties of compact stars are modeled in the framework of general
relativity, where the Einstein equations are solved for a static (non-rotating),
spherical star resulting in the Tolman--Oppenheimer--Volkoff (TOV) equations
\cite{Tolman:1939jz,Oppenheimer:1939ne,Glendenning:1997wn}
\begin{eqnarray}
\frac{dm(r)}{dr} &=& 4\pi r^{2} \varepsilon(r) \\
\frac{dp(r)}{dr} &=& - G \frac{(\varepsilon(r)+p(r))(m(r)+4\pi p(r)r^{3})}{r(r-2Gm(r))}
\label{TOV1}
\end{eqnarray}
as well as the equation for the baryon number profile
\begin{equation}
\frac{dn_{B}(r)}{dr} = 4\pi r^2 m_{N} \frac{n_{B}(r)}{\sqrt{1-2 G m(r)/r}}~.
\end{equation}
These equations are integrated from the center of the star towards its surface, with
the radius of the star $R$ being defined by the condition $p(R)=0$ and the gravitational
mass by $M=m(R)$.
In a similar manner, the baryon mass of the star is given by $M_{B}=m_{N}n_{B}(R)$, where
$m_N$ is the nucleon mass.

To complete the solution to the TOV equations the EoS is required.
It is given by the relation $p=p(\varepsilon)$ which carries information about the
microscopic ingredients of the dense nuclear matter, as mentioned before.
Thus, the above equations have to be solved simultaneously using the equation of state
under boundary conditions at the star centre ($r=0$), taken as an input.
In this way, for a given value of $\varepsilon(r=0)$ the solution of the
TOV equations are the $p(r)$ and $m(r)$ profiles and with them the parametric relationship
$M(R)$ can be obtained.

For the present study we use the scheme suggested by Alford, Han and
Prakash~\cite{Alford:2013aca} for defining the hybrid EoS (shorthand: AHP scheme),
\begin{equation}
p(\varepsilon)= p_h(\varepsilon) \Theta(\varepsilon_H-\varepsilon)
+p_h(\varepsilon_H)\Theta(\varepsilon-\varepsilon_H)
\Theta(\varepsilon_H+\Delta\varepsilon-\varepsilon)
+ p_q(\varepsilon)\Theta(\epsilon-\epsilon_H-\Delta\epsilon),
\end{equation}
where $p_h(\varepsilon)$ is a hadronic matter EoS and $p_q(\varepsilon)$ represents the
high density matter phase, here considered as deconfined quark matter with the bag model
type EoS
\begin{equation}
p_q(\varepsilon)=c^{2}_{q}\varepsilon-B.
\end{equation}
The stiffness of this EoS is given by $c^{2}_{q}$, the squared speed of sound.
The positive bag constant $B$ assures confinement, i.e. the dominance of the hadronic EoS
at low densities.
Note that a parametrization in this form by Haensel et al.~\cite{Zdunik:2012dj} describes
pretty well the superconducting NJL model derived in~\cite{Klahn:2006iw}
and systematically explored in~\cite{Klahn:2013kga}.
In the AHP scheme, the hadronic EoS is fixed and not subject to parametric variations.
However, we shall use different well-known model EoS in our study the nonrelativistic
variational EoS of Akmal et al. \cite{Akmal:1998cf} (APR) and the density-dependent relativistic
meanfield EoS of Typel and Wolter \cite{Typel:1999yq} (DD2) in its parametrization from
Ref.~\cite{Typel:2009sy}.
Both EoS come eventually with extensions due to an excluded volume for baryons, as
described for DD2 in \cite{Benic:2014jia}.
\begin{figure}[!thb]
\includegraphics[width=0.46\textwidth]{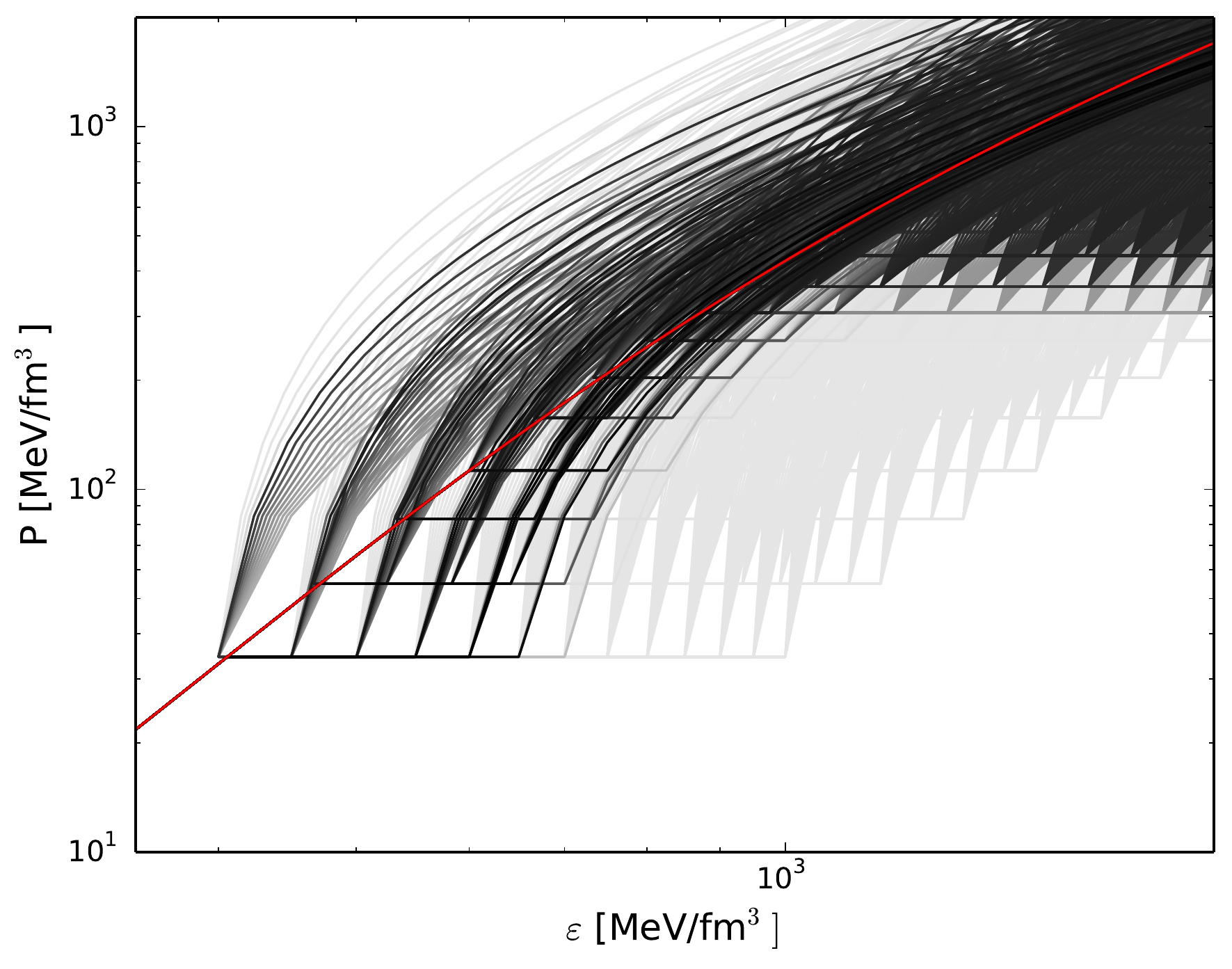}
\includegraphics[width=0.46\textwidth]{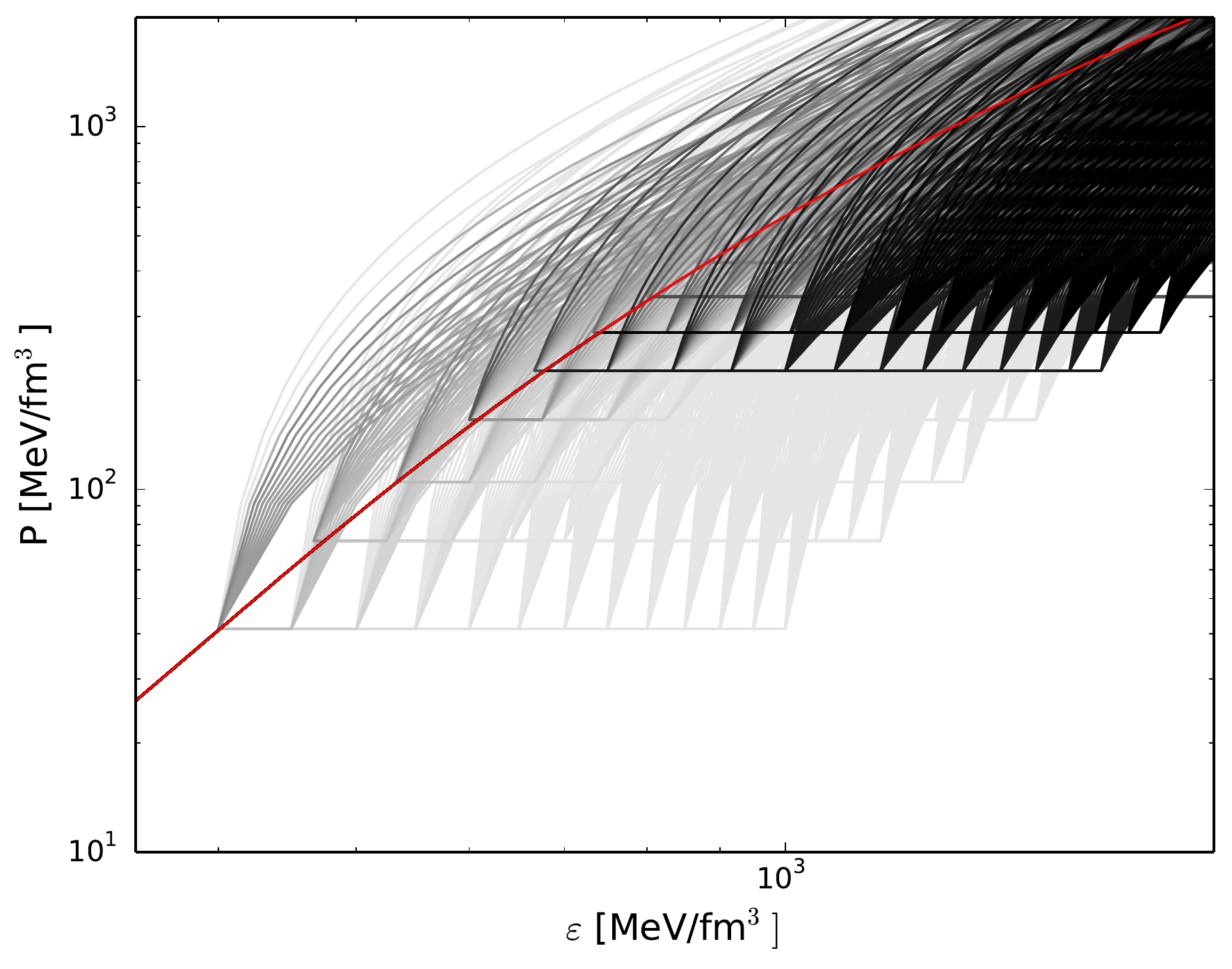}
\caption{
Varying parameters of the hybrid EoS in the AHP scheme for the
case of APR as hadronic EoS without (left) and with (right) excluded volume for baryons.
}
\label{APH_Scheme}
\end{figure}

The free parameters of the model are the transition density $\varepsilon_H$,
the energy density jump $\Delta \varepsilon \equiv \gamma \varepsilon_H$ and $c^{2}_{q}$.
The set of hybrid EoS in the plane pressure versus energy density is shown in
Fig.~\ref{APH_Scheme} for APR as the hadronic EoS.

\section{BA Formulation and Formalization}
We define the vector of free parameters
$\overrightarrow{\pi}\left(\varepsilon_c,\gamma,{c}_{q}^{2}\right)$
defining the hybrid EoS with a first order phase transition from nuclear to quark matter.

These parameters are sampled
\begin{equation}
\label{pi_vec}
\pi_i = \overrightarrow{\pi}\left(\varepsilon_k,\gamma_l,{{c}_{q}^{2}}_{m}\right),
\end{equation}
where $i = 0\dots N-1$
(here $N = N_1\times N_2\times N_3$) as $i = N_1\times N_2\times k + N_2\times l + m$ and
$k = 0\dots N_1-1$, $l = 0\dots N_2-1$, $m = 0\dots N_3-1$, here $N_1$, $N_2$ and $N_3$
denote the number of parameters for $\varepsilon_k$, $\gamma_l$ and ${{c}_{q}^{2}}_{m}$,
respectively.
Solving the TOV equations with varying boundary conditions for $\varepsilon(r=0)$ generates
a sequence of $M(R)$ curves characteristic for each of these $N$ EoS.
Subsequently, different neutron star observations with their error margins can be used to
assign a probability to each choice in from the set of EoS parameters.
We use three constraints:
(i) the mass constraint for PSR J0348+0432~\cite{Antoniadis:2013pzd},
(ii) the radius constraint for PSR J0437-4715~\cite{Bogdanov:2012md} and
(iii) the constraint on the baryon mass at the well measured gravitational mass for the
star B in the double pulsar system PSR J0737-3039~\cite{Kitaura:2006bt}, which improved the
earlier suggestion by Podsiadlowski et al.~\cite{Podsiadlowski:2005ig}.

The goal is to find the set of most probable $\pi_i$ based on given constraints using
Bayesian Analysis (BA).
For initializing BA we propose that {\it a priori} each vector of parameter $\pi_i$ has
probability equal one: $P\left(\pi_i\right) = 1$ for $\forall i$.

\subsection{Mass constraint for PSR~J0348+0432}
We propose that the error of this measurement is normal distributed
$\mathcal{N}(\mu_A,\sigma_A^2)$,
where $\mu_A = 2.01~\mathrm{M_{\odot}}$ and $\sigma_A = 0.04~\mathrm{M_{\odot}}$
are measured for PSR~J0348+0432~\cite{Antoniadis:2013pzd}.
Using this assumption we can calculate conditional probability of the event $E_{A}$
that the mass of the neutron star corresponds to this measurement
\begin{equation}
\label{p_anton}
P\left(E_{A}\left|\pi_i\right.\right) = \Phi(M_i, \mu_A, \sigma_A),
\end{equation}
where $M_i$ is the maximum mass obtained for $\pi_i$ and $\Phi(x, \mu, \sigma)$ is the
cumulative distribution function for the normal distribution.

\subsection{Radius constraint for PSR~J0437-4715}
Recently, a radius constraint for the nearest millisecond pulsar PSR~J0437-4715
have been obtained~\cite{Bogdanov:2012md} giving $\mu_B = 15.5~\mathrm{km}$ and
$\sigma_B = 1.5~\mathrm{km}$.
With these data one calculates the conditional probability of the event $E_{B}$ that
the radius of a neutron star corresponds to this measurement
\begin{equation}
\label{p_bogdan}
P\left(E_{B}\left|\pi_i\right.\right) = \Phi(R_i, \mu_B, \sigma_B).
\end{equation}

\subsection{$M$--$M_B$ Relation Constraint for PSR J0737-3039(B)}
This constraint gives a region in the $M$--$M_B$ plane.
We need to estimate the probability of a point ${\cal M}_i = \left({M}_i, {M_B}_i\right)$ to
be close to the point $\mu = \left(\mu_G, \mu_B\right)$.
The mean values $\mu_G = 1.249$, $\mu_B = 1.36$ and standard deviations
$\sigma_{M} = 0.001$, $\sigma_{M_B} = 0.002$ are given in \cite{Kitaura:2006bt}.
The probability can be calculated by following formula:
\begin{equation}
\label{p_kitaura}
P\left(E_{K}\left|\pi_i\right.\right) = \left[ \Phi\left(\xi_G\right) -
\Phi\left(-\xi_G\right) \right]\cdot\left[ \Phi\left(\xi_B\right) -
\Phi\left(-\xi_B\right) \right],
\end{equation}
where
$\Phi\left(x\right) = \Phi\left(x, 0, 1\right)$,
$\xi_G = \displaystyle\frac{\sigma_{M}}{d_{M}}$ and
$\xi_B = \displaystyle\frac{\sigma_{M_B}}{d_{M_B}}$,
$d_{M}$ and $d_{M_B}$ are absolute values of components of vector
$\mathrm{\textbf{d}} = \mathrm{\bf\mu} - \mathrm{\textbf{M}}_i$, here
$\mathrm{\bf\mu} = \left(\mu_G, \mu_B\right)^T$ was given in \cite{Kitaura:2006bt} and
$\mathrm{\textbf{M}}_i = \left({M}_i, {M_B}_i\right)^T$ stems for the solution of TOV
equations for the $i^{\mathrm{th}}$ vector of EoS parameters $\pi_i$.
Note that formula (\ref{p_kitaura}) does not correspond to a multivariate normal
distribution.

\subsection{Calculation of {\it a posteriori} Probabilities}
Note, that these measurements are independent on each other.
Therefore, the complete conditional probability of the event that a compact object
constructed with an EoS characterized by $\pi_i$ fulfils all constraints is
\begin{equation}
\label{p_event}
P\left(E\left|\pi_i\right.\right) = P\left(E_{A}\left|\pi_i\right.\right)
\times P\left(E_{B}\left|\pi_i\right.\right) \times P\left(E_{K}\left|\pi_i\right.\right).
\end{equation}
Now, we can calculate probability of $\pi_i$ using Bayes' theorem:
\begin{equation}
\label{pi_apost}
P\left(\pi_i\left|E\right.\right)=\frac{P\left(E\left|\pi_i\right.\right)P\left(\pi_i\right)}
{\sum\limits_{j=0}^{N-1}P\left(E\left|\pi_j\right.\right)P\left(\pi_j\right)}.
\end{equation}

\section{Results and Discussion}

We apply the scheme of BA for probabilistic estimation of the EoS given by vector
parameter $\vec{\pi}$.
Varying the three parameters in the intervals
$400 < \varepsilon[\mathrm{MeV/fm^{3}}] < 1200$, $0 < \gamma < 1.5$ and
$0.3 < {c}_{s}^{2} < 1.0$ with $N_1=N_2=N_3=13$ we explore a set of $N=2197$ hybrid EoS
for each choice of hadronic EoS.
\begin{figure}[!thb]
\begin{center}$
\begin{array}{cc}
\includegraphics[width=0.46\textwidth]{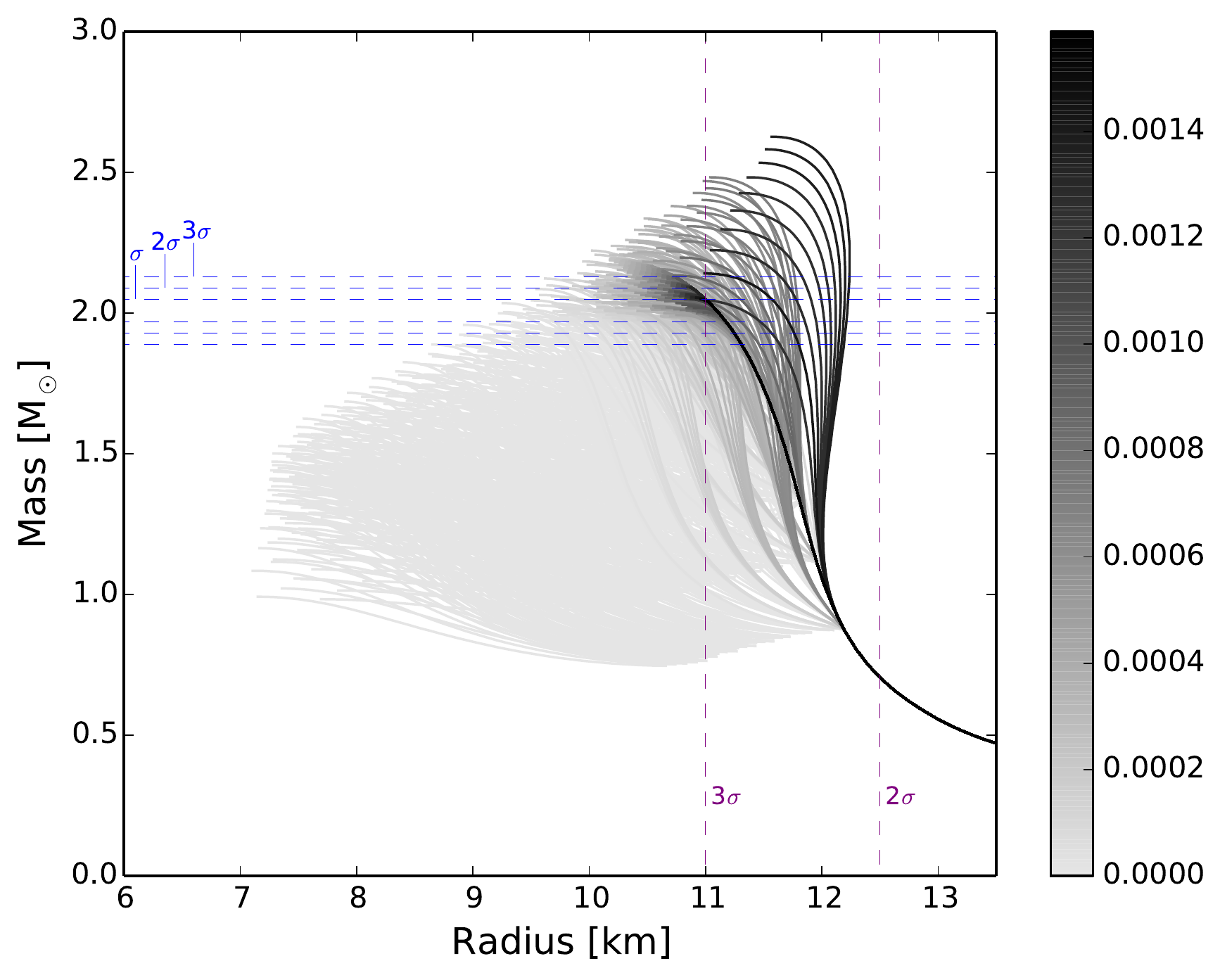}
\includegraphics[width=0.46\textwidth]{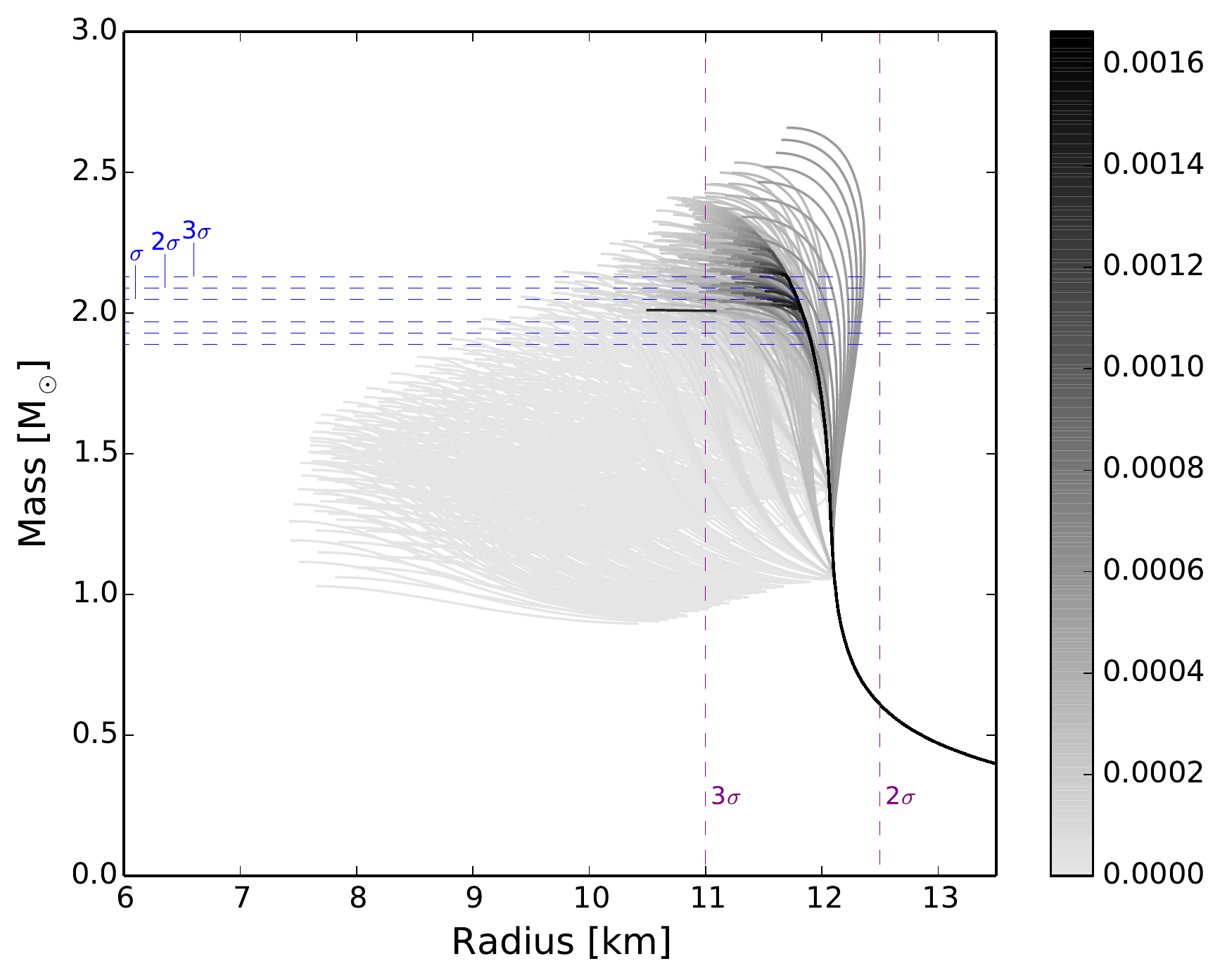}
\\
\includegraphics[width=0.46\textwidth]{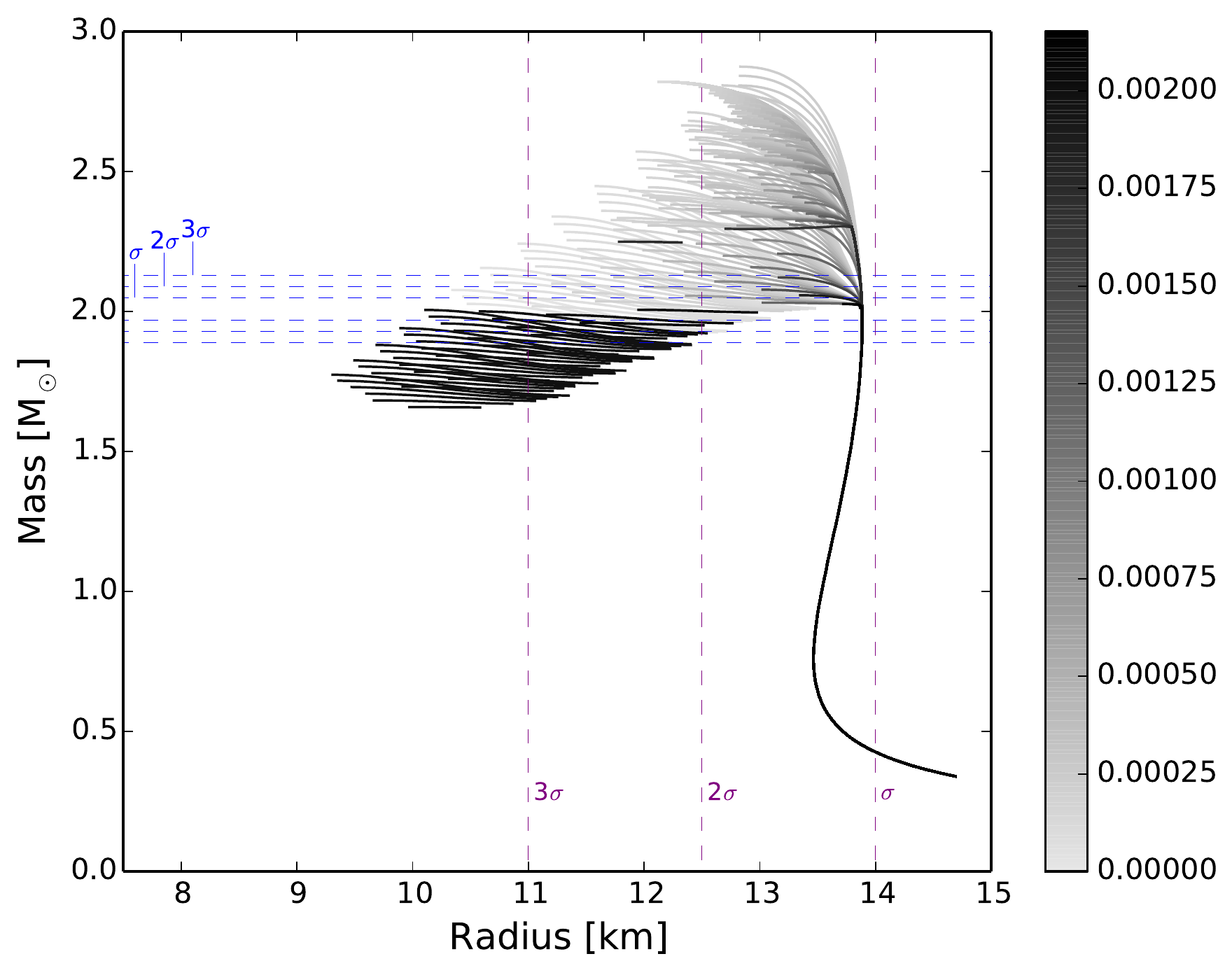}
\includegraphics[width=0.46\textwidth]{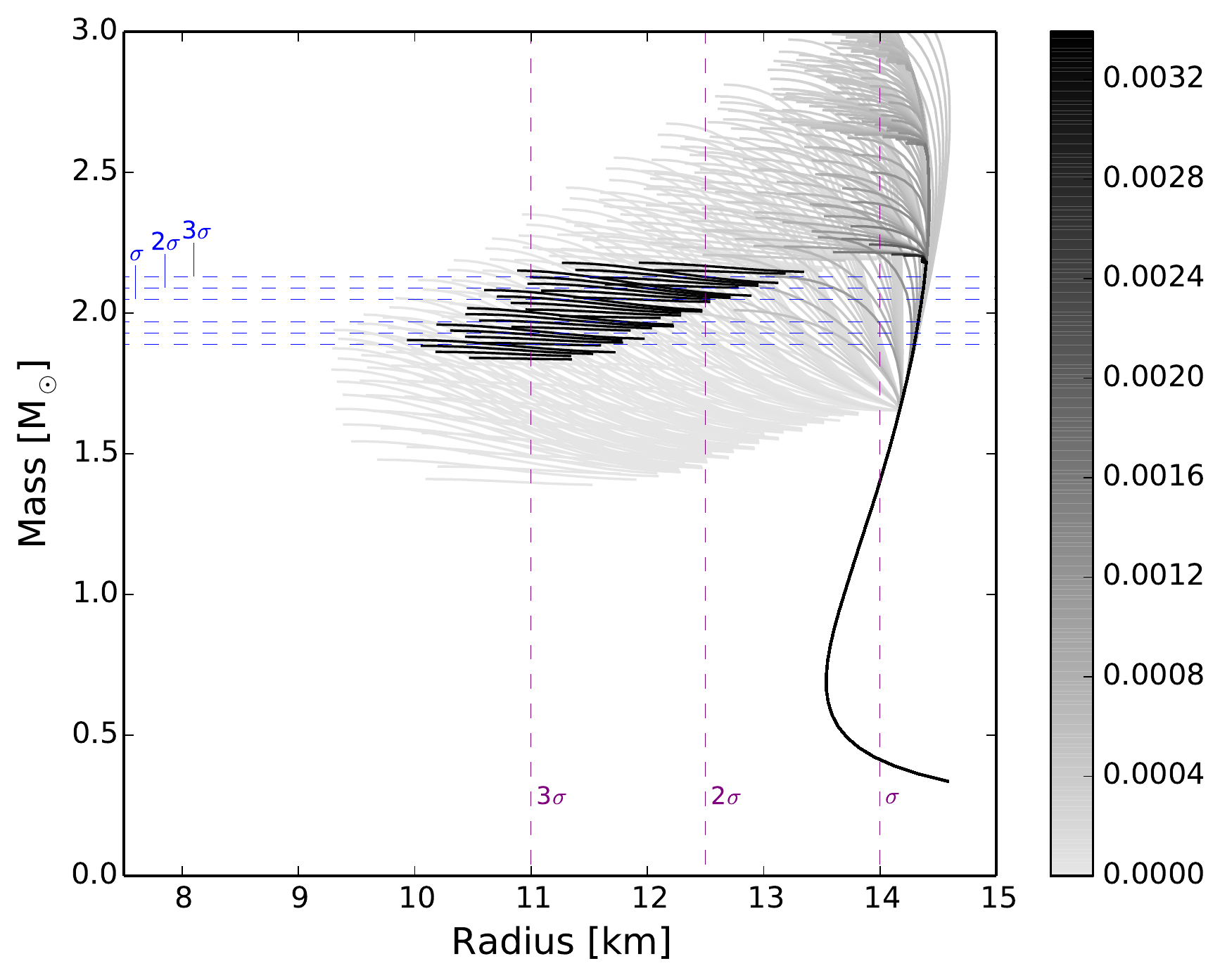}
\end{array}$
\end{center}
\caption{
\textbf{Top row:}
Mass-radius sequences for the AHP scheme with the APR EoS for baryonic matter.
Only for the hybrid EoS with baryonic excluded volume effect high mass twin star
configurations are obtained (right panel).
\textbf{Bottom row:}
Same as top row with the DD2 EoS and both cases accounting for a baryonic excluded volume,
the larger one in the right panel.
High-mass twin configurations are a rather typical feature for these stiff hadronic EoS
combined with sufficiently stiff quark matter ones.
The greyscale of the lines indicated the probability for the corresponding EoS.
}
\label{results}
\end{figure}
The results for the mass-radius sequences are presented in Fig.~\ref{results} where the
greyscale of the lines indicates the probability for the corresponding EoS parameter set
under the above constraints.
The four panels of Fig.~\ref{results} differ only in the choice of hadronic EoS.
The top row shows results for the APR-based hybrid EoS without (left) and with (right)
baryonic excluded volume effect.
Only for the latter case high mass twin stars are obtained which lie on the sequence of
the so-called "third family" of compact stars, separated from the second one (ordinary
neutron stars) by a set of unstable configurations not shown in the figure.
Hybrid EoS based on the stiffer DD2 EoS are used to obtain the sequences shown in the
bottom row, with the larger excluded volume in the right panel.
As can be seen in Fig.~\ref{results}, for the stiffer DD2-based hybrid EoS
it is quite typical to achieve a third family of compact star sequences at the high mass
of 2M$_{\odot}$.
Actually, in the BA suggested here, these twin star configurations have the highest
probability measure.
In order to support these high-mass twins, the excluded volume correction
should be introduced (see, e.g.,~\cite{Blaschke:2013rma,Blaschke:2013ana,Benic:2014jia}).

Note that the occurrence of high-mass twins is testable by observations.
It requires precise radius measurements of 2M$_{\odot}$ pulsars in order to verify
the existence of an almost horizontal branch of high-mass pulsars with almost the
same mass but radii differing by up to a few kilometers.
As a necessary condition for such sequences the hadronic star sequence should have radii
exceeding $12$ km.
The observation of this striking high-mass twin phenomenon would indicate a strong
first-order phase transition in neutron star matter which gives important constraints
for searches of a critical endpoint in the QCD phase diagram.

\section{Acknowledgment}
This research was partially supported by Bogoliubov-Infeld Program and by the 
Ter-Antonian-Smorodinsky Program for bilateral collaboration of JINR Dubna with
Universities and Institutes in Poland  and in Armenia, resp.; 
and by the COST Action MP1304 "NewCompStar". 
A.A. acknowledges JINR grant No. 14-602-01.
D.A. and D.B. were supported by the Polish Narodowe Centrum Nauki (NCN) under grant
No. UMO-2011/02/A/ST2/00306.

\end{document}